\documentclass[12pt,a4paper]{article}
\usepackage{amsmath}
\usepackage{amssymb}
\usepackage{amsthm}
\usepackage{float}
\usepackage{amsfonts}
\usepackage{graphicx}
\usepackage{verbatim}
\usepackage[left=2cm,right=2cm,top=3cm,bottom=2.5cm]{geometry}
\usepackage[numbers]{natbib}
\usepackage[utf8]{inputenc}
\usepackage[usenames,dvipsnames,svgnames]{xcolor}
\usepackage[colorlinks=true,
      linkcolor=red,
      urlcolor=gray,
      citecolor=blue]{hyperref}

\def\myalign#1{%
  \def\trule{\noalign{\smallskip\hrule\medskip}}
  \def\nebc{\nearrow\bigcup}
  \def\sebc{\searrow\bigcup}
  \def\pminf{{}_{-\infty}|^{+\infty}}
  \let\Inf\infty
  \def\amp{&} 
  \vbox{\mathsurround0pt\openup1\jot
    \halign{%
      &$\displaystyle##\hfil\tabskip0pt$&\amp##\tabskip1em\crcr
      \noalign{\hrule height1pt\smallskip}#1\noalign{\smallskip\hrule height1pt}\crcr}}}
      
\begin{document}
[DOI:10.1142/S0218271820501205]\\
\begin{center}
\textbf{On multifluid perturbations in scalar-tensor cosmology}
\end{center}
\hfill\\
Joseph Ntahompagaze$^{1}$, Shambel Sahlu$^{2,3}$, Amare Abebe$^{4}$ and Manasse R. Mbonye$^{1,5}$\\
\hfill\\
$^{1}$Department of Physics, College of Science and Technology, University of Rwanda, Rwanda\;\;\; \; \hfill\\ 
$^2$Department of Physics, College of Natural and Computational Science, Wolkite University,\;\;\; \; Ethiopia\;\;\; \hfill\\
 $^{3}$Astronomy and Astrophysics Research and Development Department, Entoto Observatory \;\;\;\;\; and Research Center,
Ethiopian Space Science and Technology Institute, Ethiopia\hfill\\ 
$^{4}$Center for Space Research, North-West University, South Africa\;\;\;\;\;\;\;\;\;\;\;\;\;\;\;\;\;\;\;\;\;\;\;\;\;\;\;\;\;\;\;\;\;\;\;\;\;\;\;\;\;\; \;\;\;  \hfill\\
$^{5}$Rwanda Academy of Sciences, Kigali, Rwanda\;\;\;\;\;\;\;\;\;\;\;\;\;\;\;\;\;\;\;\;\;\;\;\;\;\;\;\;\;\;\;\;\;\;\;\;\;\;\;\;\;\;\;\;\;\;\;\;\;\;\;\;\;\;\;\;\;\;\;\;\;\;\;\;\;\;\;\;\;\;\;\;\;\;\;\;\;\;\;\;\;\;\;\;\; \hfill\\
\hfill\\
Correspondence: ntahompagazej@gmail.com\;\;\;\;\;\;\;\;\;\;\;\;\;\;\;\;\;\;\;\;\;\;\;\;\;\;\;\;\;\;\;\;\;\;\;\;\;\;\;\;\;\;\;\;\;\;\;\;\;\;\;\;\;\;\;\;\;\;\;\;\;\;\;\;\;\;\;\;\;\;\;\;\;\;\;
\begin{center}
\textbf{Abstract}
\end{center}
In this paper the scalar-tensor theory is applied to the study of perturbations in a multi-fluid universe, using the 1+3 covariant approach. Both scalar and harmonic decompositions are instituted on the perturbation equations. In particular, as an application, we study perturbations on a background FRW cosmology consisting of both radiation and dust in the presence of a scalar field. We consider both radiation-dominated and dust-dominated epochs, respectively, and study the results. During the analysis, quasi-static approximation is instituted. It is observed that the fluctuations of the energy density decrease with increasing redshift, for different values of $n$ of a power law $R^{n}$ model.\\
\hfill\\
\textit{keywords:} $f(R)$ gravity --- scalar-tensor --- scalar field ---cosmology ---covariant perturbation

\section{Introduction}
Models to explain cosmic large-scale structure formation have usually employed a perturbative approach.
Linear perturbations have severally been discussed in General Relativity (GR) to studies of various epochs using several techniques, including applications of 
the $1+3$ covariant formalism, applied to studies of various cosmic epochs \cite{ellis1989covariant,bruni1992cosmological}.  
On the other hand, similar studies at linear order have lately been done in modified gravity theories such as $f(R)$ gravity \cite{Ananda2,amare2,27}. 
In a recent paper 
\cite{ntahompagaze2017f} (referred to as Paper I of this series), we have discussed $f(R)$ gravity in scalar-tensor (ST) theories, where the equivalence 
between $f(R)$ theory and ST theory was explored. We note that in ST theory, the $1+3$ covariant formalism has previously been applied to 
study linear perturbations in the vacuum case \cite{SanteCarloni4, osano2007gravitational}. In \cite{ntahompagaze2017study} (referred to as Paper II of this series), we have studied perturbations of a two-fluid system 
at linear order.
In the current paper this equivalence is
extended to discuss linear perturbations for a multi-fluid system in ST theory. Here the scalar 
field is considered as one of the fluids. We utilize the $1+3$ covariant formalism to study perturbations in cosmology. 
We consider our current discussion of perturbations of the multi-fluid system, as a logical extension of our work in this series. 

\noindent The $1+3$ linear covariant perturbation in a multi-fluid system was first studied in GR \cite{dunsby1992covariant}.
This is motivated by the observation that the physical universe is composed of many fluids
say, relativistic particles,
radiation, dust, cold dark matter ($CDM$) and others.
A further step has been made where the extension to multi-fluid system studies is 
taken into account in modified theories of gravity, say $f(R)$ models \cite{amare4}. In this paper, we treat the behavior of density perturbations in power law, $R^{n}$, models. These models were first proposed in \cite{Buchdahl1}. Later, they were explored in several works, see for example in \cite{barrow1983stability} where stability analyses of different $f(R)$ models are treated. 
The $f(R)$ theory and other modified gravity theories have received a significant attention after the observations of the cosmic acceleration \cite{1,perlmutter2003measuring}. The motivation for this interest is that cosmic acceleration epoch can be reproduced without the implication of dark energy hypothesis \cite{nojiri2007introduction}.
Among the theories treated linear perturbations using $1+3$ covariant approach, one can name the recent work done in \cite{sahlu2020scalar}, where the treatment was about torsion gravity theory with interest in two fluid systems. 
 
\noindent In this work, we use the equivalence between $f(R)$ and ST theory first developed in our previous work \cite{ntahompagaze2017f} (referred to as Paper I of this series) to study perturbations in a multifluid cosmology obeying a power law, $R^{n}$. This is a natural extension to the work we have previously done in this series (referred to as Paper II) in which linear perturbations were applied separately to a radiation system and a dust system both in the presence of the scalar field. In particular, we apply the study to an FRW cosmology with radiation and-or dust in the presence of the scalar field. We analyze the evolution of such a universe with a focus to short-wavelength perturbation modes.

\noindent The following is the organization of the paper. In Section \ref{mathtool}, we provide the $1+3$ covariant approach in ST theory together with the definition of useful covariant gradient variables for the total and component fluids
followed by both the evolution equations and harmonic decomposition. 
 We analyze evolutions of the perturbations for the short-wavelength regime in Section \ref{RADIATIONDUSTUNIVERSE}. In Section \ref{conclusion}, we provide conclusion.

\noindent The adopted spacetime signature is $(- + + +)$ and unless stated otherwise, we
have used the convention $8\pi G \equiv c \equiv 1$, where $G$ is the gravitational constant and
$c$ is the speed of light.

\section{Mathematical Tools}\label{mathtool}
\subsection{The 1+3 covariant approach for scalar-tensor theories}\label{1+3COV}
This approach is the way of dividing the space-time into foliated spacelike hypersurfaces and a perpendicular 4-vector-field.
In this process, the cosmological manifold $(\mathcal{M},g)$ is decomposed into the sub-manifold $(M,h)$ which has a 
perpendicular 4-velocity field vector $u^{a}$. In this study, the background under consideration is the FLRW spacetime.
The 4-velocity field vector $u^{a}$ is defined as
\begin{equation}
u^{a}=\frac{dx^{a}}{d \tau}\; , \text{ such that } u_{a}u^{a}=-1\; .
\end{equation} 
This approach helps in the decomposition of the metric $g_{ab}$ into the projection tensor $h_{ab}$ as \cite{3,amare4,amare2}:
\begin{equation}
g_{ab}= h_{ab}-u_{a}u_{b}\; .
\end{equation} 
The covariant time derivative  and spatial covariant derivative $\tilde{\nabla}$ for a given tensor $T^{ab}_{cd}$ are given as
\begin{equation}
\dot{T}^{ab}_{cd}=u^{e}\nabla_{e}T^{ab}_{cd} \; , \quad
\tilde{\nabla}_{e}T^{ab}_{cd}=h^{a}_{f}h^{b}_{g}h^{p}_{c}h^{q}_{d}h^{r}_{e}\nabla_{r}T^{fg}_{pq}\; .
\end{equation} 
The matter energy-momentum tensor $T_{ab}$ is also decomposed with the 1+3 covariant approach and it is given as \cite{dunsby1992covariant,dunsby1991gauge,3,amare2}:
\begin{equation}
T_{ab}=\mu u_{a}u_{b}+q_{a}u_{b}+u_{a}q_{b}+ph_{ab}+\pi_{ab}\; ,\label{13}
\end{equation}
where $\mu$, $q^{a}$, $p$ and $\pi_{ab}$  are the relativistic energy density,
momentum density,  isotropic pressure, and trace-free anisotropic pressure of the fluid respectively.
For a perfect fluid, $q^{a}=\pi_{ab}=0$.
In this approach, the kinematic quantities which are obtained from irreducible parts of the decomposed $\nabla_{a}u_{b}$ are given as \cite{dunsby1992covariant,dunsby1991gauge,3}
\begin{equation}
\nabla_{a}u_{b}=\tilde{\nabla}_{a}u_{b}-u_{a}\dot{u}_{b}=\frac{1}{3}\theta h_{ab}+\sigma_{ab}+\omega_{ab}-u_{a}\dot{u}_{b}\; , \label{definitionofu}
\end{equation}
where the volume rate of expansion of the fluid $\theta =\tilde{\nabla}_{a}u^{a}$, the Hubble parameter $H=\frac{\theta}{3}=\frac{\dot{a}}{a}$, 
the rate of distortion of the matter flow $\sigma_{ab}=\tilde{\nabla}_{<a}u_{b>}$.
The vorticity tensor $\omega_{ab}=\tilde{\nabla}_{[a}u_{b]}$ is the skew-symmetric vorticity tensor and describes the rotation of the matter 
relative to a non-rotating frame.
The relativistic acceleration vector (not that of the expansion of the universe) is given as
$\dot{u}^{a}=u^{b}\nabla_{b}u^{a}$. These kinematic quantities provide informations about the overall spacetime kinematics. 

\subsection{The $f(R)$ theory in scalar-tensor language}\label{STlanguage}
Let us consider the action that represents $f(R)$ gravity given as
\begin{equation}\label{fRaction1}
I=\frac{1}{2\kappa}\int d^{4}x\sqrt{-g}\left[f(R)+\mathcal{L}_{m}\right]\; ,
\end{equation}
where $\kappa= 8\pi G$, $R$ is Ricci scalar and $\mathcal{L}_{m}$ is the matter Lagrangian.
The above action for $f(R)$ theory of gravity can have its equivalence in ST theory as far as one has the appropriate definition of 
the scalar field $\phi$. The action related to this concern is presented  as \cite{7, Wands1994}
\begin{equation}
I=\frac{1}{2\kappa}\int d^{4}x\sqrt{-g}\left[\phi R-V(\phi)+\mathcal{L}_{m}\right]\;,
\end{equation}
where $V(\phi)$ is the potential.
 We therefore define the scalar field as \cite{scalar5}
\begin{equation}
\phi=f'-1\; ,
\end{equation}
where $f'=\frac{df}{dR}$ and the scalar field $\phi$ has to be invertible \cite{scalar1,7,scalar3}. From this construction
of the equivalence, one can rewrite the above action as \cite{scalar5,7}
\begin{equation}
I_{f(\phi)}=\frac{1}{2\kappa}\int d^{4}\sqrt{-g}\left[f(\phi)+\mathcal{L}_{m}\right]\; ,\label{frstt1}
\end{equation}
where $f(\phi)$ is the function of $\phi(R)$. 
The following evolution equations are equivalent to the field equations that one can have from the action in Eq. \eqref{frstt1} once
a variation with respect to the metric $g_{\mu\nu}$ is performed. The Friedmann and the Raychaudhuri equations are given as 
\begin{eqnarray}
&&\theta^{2}=3(\tilde{\mu}_{m}+\mu_{\phi})-\frac{9K}{a^{2}}\; ,\label{fried}\\
&&\dot{\theta}+\frac{1}{3}\theta^{2}+\frac{1+3\omega}{2}\tilde{\mu}_{m}+\frac{1}{2}(\mu_{\phi}+3p_{\phi})=0\; ,\label{duri}
\end{eqnarray}
respectively,
 where $w$ is barotropic equation of state (EoS) parameter, where $p_{m}=w\mu_{m}$ , $\mu_{\phi}$ is the energy density and $p_{\phi}$ is isotropic pressure of
the scalar fluid respectively; $K$ stands for curvature and has the values $0,-1,+1$ for flat, open and closed universe respectively,
$a$ is the scale factor and $\tilde{\mu}_{m}=\frac{\mu_{m}}{\phi+1}$ with $\mu_{m}$ being the matter energy density.
Now since the background is considered to be the FLRW spacetime, the background quantities: energy density 
and isotropic pressure for the curvature 
fluid are considered as 
\begin{eqnarray}
&&\mu_{\phi}=\frac{1}{\phi+1}\left[\frac{1}{2}\Big((\phi+1)R-f\Big)-\theta\dot{\phi}\right]\; ,\label{14aa}\\
&&p_{\phi}= \frac{1}{\phi+1}\left[\frac{1}{2}\Big(f-R(\phi+1)\Big)+\ddot{\phi}-\frac{\dot{\phi}\dot{\phi}'}{\phi'}\phi''
+\frac{\phi''\dot{\phi}^{2}}{\phi'^{2}}+\frac{2}{3}\theta \dot{\phi}\right]\; .\label{14bb}
\end{eqnarray}
For FLRW background spacetimes, we have the quantities defined in Eq. \eqref{definitionofu} behaving as
\begin{equation}
 \sigma_{ab}=0\; , \quad \omega_{a}=0=\dot{u}_{a}=\tilde{\nabla}_{a}\theta\; .
 \label{kinematicquantitybackg}
\end{equation}
Also for any scalar quantity $f$, in the background we have
\begin{equation}
 \tilde{\nabla}_{a}f=0,
\end{equation}
and hence
\begin{equation}
\tilde{\nabla}_{a}\mu=0=\tilde{\nabla}_{a}p=q^{\phi}_{a}=\tilde{\nabla}_{a}\phi\; , \quad =\pi^{\phi}_{ab}=0\; . \label{nablapsi}
\end{equation}
The energy conservation or simply the continuity equations for matter and curvature fluid are given as
\begin{eqnarray}
&&\dot{\mu}_{m}=-\theta(\mu_{m}+p_{m})\; , \\
&&\dot{\mu}_{\phi}=-\theta(\mu_{\phi}+p_{\phi})+\frac{\dot{\phi}\mu_{m}}{(\phi+1)^{2}}\; . \label{yyayaya}
\end{eqnarray}

\subsection{Definition of gradient variables}\label{GRADIENTVARABLESS}
The gauge-invariant quantities are key for the cosmological perturbations analysis. A gauge-invariant (GI) quantity is first order 
if it vanishes in the background. One can obtain the detail of this statement in the Stewart-Walker Lemma
\cite{stewart1974perturbations}.
We define GI quantities in the 1+3 covariant perturbations in the following way.
The GI variable that characterizes energy density perturbation in spatial variations
is given as \cite{bruni1992gauge}
\begin{equation}
D_{a}=\frac{a}{\mu}\tilde{\nabla}_{a}\mu\; . 
\end{equation}
The ratio $\frac{a}{\mu}$ helps to evaluate the magnitude of energy density perturbations relative to the background \cite{SanteCarloni4}. 
We also define the other two quantities. The spatial gradient of the volume expansion \cite{dunsby1991gauge}
\begin{equation}
Z_{a}=a\tilde{\nabla}_{a}\theta\; , \label{Za} 
\end{equation}
and the spatial gradient of the 3-Ricci scalar $\tilde{R}$ as
\begin{equation}
C_{a}=a^{3}\tilde{\nabla}_{a}\tilde{R}\; .\label{C_{a1}} 
\end{equation}
We also define two other gradient variables $\Phi_{a}$ and $\Psi_{a}$ that characterize perturbations due to scalar field and momentum of
scalar field as \cite{ntahompagaze2017study}
\begin{equation}
 \Phi_{a}=a\tilde{\nabla}_{a}\phi\; ,
\end{equation}
\begin{equation}
\Psi_{a} =a\tilde{\nabla}_{a}\dot{\phi}\; .
\end{equation}
Since the universe is composed of different fluids (radiation, CDM, etc), it is useful to define the gradient variables
that are consistent with multi-fluid 1+3 covariant perturbations. First of all we rewrite the total matter energy-momentum tensor defined in Eq. \eqref{13}
in the following form \cite{dunsby1992covariant,amare4}:
\begin{equation}
T^{m}_{ab}=\sum_{i} T^{i}_{ab}\; , \label{multifluid1}
\end{equation}
where $T^{i}_{ab}$ represents the energy-momentum tensor of the i$^{th}$ fluid component and it is defined as
\begin{equation}
T^{i}_{ab}=\mu_{i} u^{i}_{a}u^{i}_{b}+q^{i}_{a}u^{i}_{b}+u^{i}_{a}q^{i}_{b}+p_{i}h^{i}_{ab}+\pi^{i}_{ab}\; ,\label{multifluid2}
\end{equation}
with the definition that
\begin{equation}
h^{i}_{ab}=g_{ab}+u^{i}_{a}u^{i}_{b}\; . 
\end{equation}
From now on, the $i$ and $j$ as superscript (or subscript) will be indicating the $i^{\text{th}}$ and $ij^{\text{th}}$ fluid components respectively not the running index.  
One can define the relative velocity of 
the i$^{th}$ component with respect to the observer as \cite{dunsby1992covariant,amare4}
 \begin{equation}
  V^{i}_{a}\equiv u^{i}_{a}-u_{a}\; ,\label{relativevelocitVi}
 \end{equation}
where $V^{i}_{a}=0$ for homogeneous medium and $V^{i}_{a}\neq 0$ for inhomogeneous medium.
We want to define the vector gradient variables for the above described multifluid component  \cite{dunsby1992covariant,amare4}.
The spatial gradient will be defined as
 \begin{equation}
D^{i}_{a}=\frac{a}{\mu_{i}}\tilde{\nabla}_{a}\mu_{i}\; . 
\end{equation}
The gradient for pressure will be
\begin{equation}
Y^{i}_{a}= \tilde{\nabla}_{a}p^{i}\; ,
\end{equation}
The gradient variable for entropy density is given as
\begin{equation}
 \varepsilon^{i}_{a}=\frac{a}{p_{i}}\left(\frac{\partial p^{i}}{\partial s_{i}}\right)\tilde{\nabla}_{a}s_{i}\; ,
\end{equation} 
$s_{i}$ being the entropy for the $i^{th}$ component.
Then, we can write gradient variable for entropy perturbation $\varepsilon_{a}$ of total fluid such that
\begin{equation}
p_{m}\varepsilon_{a}=\sum_{i}p_{i}\varepsilon^{i}_{a}+\frac{1}{2}\sum_{i,j}\frac{h_{i}h_{j}}{h}(c^{2}_{si}-c^{2}_{sj})S^{ij}_{a}\; ; 
\label{varepsilonmatter1}
\end{equation}
where $h=\mu_{m}+p_{m}$ and $h_{i}=\mu_{i}+p_{i}$, here the subscript $m$ is not  a running index, it only indicates matter.
For the treatment of the adiabatic and isothermal perturbations the following relative gradient varibles are more important:
\begin{eqnarray}
&S^{ij}_{a}\equiv \frac{\mu_{i}D^{i}_{a}}{(\mu_{i}+p_{i})}-\frac{\mu_{j}D^{j}_{a}}{(\mu_{j}+p_{j})}\;,\\ 
&V^{ij}_{a}\equiv V^{i}_{a}-V^{j}_{a}\; . 
\end{eqnarray}
Again, here the $ij$ present in the above equations does not make the $LHS$ tensors, since $ij$ are not running indices as stated earlier.
The scalar perturbations are believed to be the ones responsible for large-scale structure formation.
We extract the scalar part from the quantities under consideration using the local decomposition for a quantity $X_{a}$ as \cite{amare4,27}
\begin{equation}
a\tilde{\nabla}_{b}X_{a}=X_{ab}=\frac{1}{3}h_{ab}X+\Sigma^{X}_{ab}+X_{[ab]}\; , 
\end{equation}
where $\Sigma^{X}_{ab}=X_{(ab)}-\frac{1}{3}h_{ab}X$ describes shear and $X_{[ab]}$ describes vorticity. 
Now let us apply the comoving differential operator $a\tilde{\nabla}_{a}$ to $D_{a},Z_{a}$ and $C_{a}$ to have
\begin{equation}
\Delta_{m}=a\tilde{\nabla}^{a}D^{m}_{a}\; , Z=a\tilde{\nabla}^{a}Z_{a}\; , 
  C=a\tilde{\nabla}^{a}C_{a}\; ,
\Psi =a\tilde{\nabla}^{a}\Psi_{a}\; , 
\text{ and }
\Phi=a\tilde{\nabla}^{a}\Phi_{a}\; . 
\end{equation}
Note that the above variables are Gauge-invariants as $D_{a},Z_{a},C_{a}, \Phi_{a}$ and $\Psi_{a}$ are Gauge-invariants.
For multifluid component, we have the scalar gradient variables as
 \begin{equation}
\Delta_{i}=a\tilde{\nabla}^{a}D^{i}_{a}\; ,
 \varepsilon_{i}=a\tilde{\nabla}^{a}\varepsilon^{i}_{a}\; ,
S_{ij}=a\tilde{\nabla}^{a}S^{ij}_{a}\; , 
\text{ and }
V_{ij}=a\tilde{\nabla}^{a}V^{ij}_{a}\; . 
\end{equation}

\subsection{Linear evolution equations}\label{LinearEqsection}
In this section, we are using linear quantities for both scalar field fluid energy density and pressure given as
\begin{equation}
 \mu_{\phi}=\frac{1}{\phi+1}\left[\frac{1}{2}\Big((\phi+1)R-f\Big)-\theta\dot{\phi}-\phi'\tilde{\nabla}^{2}R\right]\; ,
\end{equation}
and 
\begin{equation}
p_{\phi}=\frac{1}{\phi+1}\left[\frac{1}{2}\Big(f-R(\phi+1)\Big)+\ddot{\phi}-\frac{\dot{\phi}\dot{\phi}'}{\phi'}\phi''
+\frac{\phi''\dot{\phi}^{2}}{\phi'^{2}}+\frac{2}{3}\theta \dot{\phi}-\frac{2\phi'\tilde{\nabla}^{2}R}{3}\right]\; . 
\end{equation}
Note that these two equations are the extension of Eqs. \eqref{14aa} and \eqref{14bb} with first-order contributions included.
Here, we provide scalar evolution equations for perturbations of quantities with irrotational fluid assumption. This means that we will make the term
$\omega_{a}=0$ for the rest of the work. For the total matter fluid one has scalar perturbation equations as \cite{SanteCarloni1,amare4,27}
\begin{eqnarray}
&&\dot{\Delta}_{m}=-(1+w)Z+w\theta \Delta_{m}\; . \label{dotDeltammulti1}
\end{eqnarray}
The evolution for comoving volume expansion is given as
\begin{eqnarray}
\begin{split}
&\dot{Z}=\Big(\frac{\dot{\phi}}{\phi+1}-\frac{2\theta}{3}\Big)Z-\frac{c^{2}_{s}}{(1+w)}\tilde{\nabla}^{2}\Delta_{m}
-\frac{w}{(1+w)}\tilde{\nabla}^{2}\varepsilon
-\frac{1}{\phi+1}\tilde{\nabla}^{2}\Phi\\
&+\Big[\frac{1}{2\phi'}+\frac{2\mu_{m}-f-2\theta\dot{\phi}}{2(\phi+1)^{2}}-\frac{\phi''\tilde{\nabla}^{2}R}{\phi'(\phi+1)}
+\frac{\phi'\tilde{\nabla}^{2}R}{(\phi+1)^{2}}-\frac{2K}{(\phi+1)a^{2}}\Big]\Phi\\
&+\Big[\frac{(\phi+1)c^{2}_{s}-w-1}{(w+1)(\phi+1)}\mu_{m}-\frac{c^{2}_{s}}{1+w}\Big(-\frac{1}{3}\theta^{2}+\frac{f}{2(\phi+1)}+\frac{\theta\dot{\phi}}{\phi+1}
-\frac{\phi'}{\phi+1}\tilde{\nabla}^{2}R\Big)\Big]\Delta_{m}\\
&-\frac{w}{1+w}\Big[-\frac{1}{3}\theta^{2}-\frac{\mu_{m}}{\phi+1}+\frac{f}{2(\phi+1)}+\frac{\theta\dot{\phi}}{\phi+1}
-\frac{\phi'}{\phi+1}\tilde{\nabla}^{2}R\Big]\varepsilon+\frac{\theta}{\phi+1}\Psi\; .\label{dotZmulti1}
\end{split}
\end{eqnarray}
The evolution for fluctuations in scalar field is given
\begin{eqnarray}
&&\dot{\Phi}=\Psi-\frac{c^{2}_{s}\dot{\phi}}{w+1}\Delta_{m}-\frac{w\dot{\phi}}{w+1}\varepsilon\; .\label{dotPhimulti1}
\end{eqnarray}
The evolution for fluctuations in momentum of scalar field is given
\begin{eqnarray}
&& \dot{\Psi}=\frac{\ddot{\phi}'}{\phi'}\Phi-\frac{c^{2}_{s}\ddot{\phi}}{(w+1)}\Delta_{m}
 -\frac{w\ddot{\phi}}{(w+1)}\varepsilon\; ,\label{dotPsimulti1}
\end{eqnarray}
and the above four evolution equations satisfy the constraint
\begin{equation}
\begin{split}
\frac{C}{a^{2}}&=-\Big(\frac{4}{3}\theta+\frac{2\ddot{\phi}}{\phi+1}\Big)Z+2\frac{\mu_{m}}{\phi+1}\Delta_{m}
+\Big(-2\frac{\mu_{m}}{(\phi+1)^{2}}-\frac{2\theta\ddot{\phi}'}{\phi'(\phi+1)}
+\frac{2\theta\ddot{\phi}}{\phi+1}\\
&+\frac{4K}{a^{2}(\phi+1)}+\frac{2\phi''\tilde{\nabla}^{2}R}{(\phi+1)\phi'}
-\frac{2\phi'\tilde{\nabla}^{2}R}{(\phi+1)^{2}}+\frac{f}{(\phi+1)^{2}}\Big)\Phi
+\frac{2}{\phi+1}\tilde{\nabla}^{2}\Phi\; .\label{Cmulti1}
\end{split}
\end{equation}
The scalar perturbations for component fluid are given as
\begin{eqnarray}
 &&
 \begin{split}
 \dot{\Delta}_{i}&=-(1+w_{i})Z-\theta(w_{i}-c^{2}_{si})\Delta_{i}-(1+w_{i})a\tilde{\nabla}_{b}\tilde{\nabla}^{b}V_{i}
-\frac{\theta h_{i}}{\mu_{i}h}\Big(c^{2}_{s}\mu \Delta+p\varepsilon\Big)\; ,
\end{split}\label{dotDeltaimulti1}\\
&&\dot{V}_{i}-(3c^{2}_{si}-1)\frac{\theta}{3}V_{i}=\frac{1}{ahh_{i}}\Big(-hc^{2}_{si}\mu_{i}\Delta_{i}
+h_{i}c^{2}_{s}\mu \Delta+h_{i}p\varepsilon\Big)\; , \label{dotVimulti1}\\
&&
\begin{split}
 \dot{V}_{ij}&=-(c^{2}_{si}-c^{2}_{sj})\theta V^{i} -(3c^{2}_{sj}-1)\frac{\theta}{3}V_{ij}
 -\frac{(c^{2}_{si}-c^{2}_{sj})}{a(1+w_{i})}\Delta_{i}-\frac{c^{2}_{sj}}{a}S_{ij}\; , \label{dotVijmulti1}
\end{split}\\
&& \dot{S}_{ij}=-a\theta\tilde{\nabla}^{2}V_{ij}\; ,\label{dotSilmulti1}
\end{eqnarray}
where $\tilde{\nabla}^{2}=\tilde{\nabla}_{b}\tilde{\nabla}^{b}$.
 \subsection{Harmonic decomposition}\label{HARMONIC1}
This method of harmonic decomposition has been used for 1+3 covariant linear
perturbations extensively in \cite{amare2, amare4,SanteCarloni4}. This approach 
allows one to treat the scalar perturbations equations as ordinary differential equations at each mode $k$ separately. Therefore, the analysis
becomes easier when dealing with ordinary differential equations rather than the former (partial differential) equations.
We consider the differential equation given as \cite{amare2}:
\begin{equation}
\ddot{X}+\mathcal{A}_{1}\dot{X}+\mathcal{A}_{2}X=\mathcal{A}_{3}(Y,\dot{Y})\; ,\label{harmonic1} 
\end{equation}
where $\mathcal{A}_{1},\mathcal{A}_{2}$ and $\mathcal{A}_{3}$ are independent of $X$ and they represent damping, restoring and source terms respectively. The separation
of variables for solutions of Eq. \eqref{harmonic1} is done such that $X$ and $Y$ have separable component $Q(\vec{x})$ depends on spatial 
variable $\vec{x}$ only, and $U(t)$ and $W(t)$ depend on time variable $t$ only so that
\begin{eqnarray}
X=\sum_{k}U^{k}(t)Q_{k}(\vec{x}),   \text{  and  } Y=\sum_{k}W^{k}(t)Q_{k}(\vec{x})\; , 
\end{eqnarray}
where $Q_{k}$ are the eigenfunctions of the covariant Laplace-Beltrami operator such that
\begin{eqnarray}
 \tilde{\nabla}^{2}Q=-\frac{k^{2}}{a^{2}}Q\; ,
\end{eqnarray}
and the order of harmonic $k$ (wavenumber) is 
\begin{eqnarray}
 k=\frac{2\pi a}{\lambda}\; , \label{wavenumberk}
\end{eqnarray}
where $\lambda$ is the physical wavelength of the mode. The eigenfunctions $Q$ are time independent, that means 
$\dot{Q}(x)=0$. 
\section{Radiation-dust universe}\label{RADIATIONDUSTUNIVERSE}
\subsection{Basics of the radiation-dust mixture }
To reduce the above developed multi-fluid perturbation equations, we consider a universe filled with non-interacting radiation and dust together with
scalar field. The three form the total fluid. If one assumes the flat homogeneous and isotropic universe as background (FLRW with $K=0$),
we can write down the evolution equation for radiation energy density $\mu_{r}$ and dust energy density $\mu_{d}$ as \cite{amare4}:
\begin{equation}
\dot{\mu}_{r}=-\frac{4}{3}\theta\mu_{r}\; , 
\end{equation}
and 
\begin{equation}
\dot{\mu}_{d}=-\theta\mu_{d}, 
\end{equation}
where the equation of state parameter for dust is considered to be $w_{d}=0$ and that of radiation is $w_{r}=\frac{1}{3}$.
In some equations, like in the above equations, the $r$ and $d$ superscripts (sometimes are used as subscripts) are not running indices, they only represent radiation or dust respectively.
 With this in mind,
the equation of state parameter of the total matter fluid (excludes the scalar field) is given as
\begin{equation}
 w=\frac{p_{m}}{\mu_{m}}=\frac{\mu_{r}}{3(\mu_{d}+\mu_{r})}\; .
\end{equation}
The adiabatic speed of sound in this total matter fluid (dust and radiation) is given as
\begin{equation}
c^{2}_{s}=\frac{\dot{p}_{m}}{\dot{\mu}_{m}}=\frac{4\mu_{r}}{3(3\mu_{d}+4\mu_{r})}\; . 
\end{equation}
We can also define a parameter $c^{2}_{z}$ which connects two speeds of sound $c^{2}_{sd}$ and $c^{2}_{sr}$ such that
\begin{equation}
c^{2}_{z}=\frac{1}{h}\Big(h_{r}c^{2}_{sd}+h_{d}c^{2}_{sr}\Big)=\frac{\mu_{d}}{4\mu_{r}+3\mu_{d}}\; . 
\end{equation}
We revisit Eq. \eqref{varepsilonmatter1} and assume no interaction between the two fluids under consideration and also that
$\varepsilon^{i}\approx 0$, so that we can write the entropy perturbation
\begin{equation}
p_{m}\varepsilon_{a}=\frac{1}{2}\Big(\frac{h_{d}h_{r}}{h}(c^{2}_{sd}-c^{2}_{sr})S^{dr}_{a}+\frac{h_{r}h_{d}}{h}(c^{2}_{sd}
-c^{2}_{sd})S^{rd}_{a}\Big)\; . 
\end{equation}
Using the definition of $S^{ij}_{a}$, we can write
\begin{equation}
S^{dr}_{a}=-S^{rd}_{a}\; .
\end{equation}
Therefore, our leading equation becomes
\begin{equation}
p_{m}\varepsilon_{a}=-\frac{4\mu_{d}\mu_{r}}{3(3\mu_{d}+4\mu_{r})}(c^{2}_{sd}-c^{2}_{sr})S^{dr}_{a}\; . 
\end{equation}
Dividing both sides of the above equation by $p_{m}$, one has
\begin{equation}
\varepsilon_{a}=-\frac{4\mu_{d}}{(3\mu_{d}+4\mu_{r})}(c^{2}_{sd}-c^{2}_{sr})S^{dr}_{a}\; .  
\end{equation}
Its scalar equation is given as
\begin{equation}\label{varepsilonnnnnnnnnnn}
\varepsilon=-\frac{4\mu_{d}}{(3\mu_{d}+4\mu_{r})}(c^{2}_{sd}-c^{2}_{sr})S_{dr}\; , 
\end{equation}
whereas the harmonically decomposed quantity is given as
\begin{equation}
\varepsilon^{k}=-\frac{4\mu_{d}}{(3\mu_{d}+4\mu_{r})}(c^{2}_{sd}-c^{2}_{sr})S^{k}_{dr}\; .\label{scalarvarepsilon1}  
\end{equation}
Before writing down the evolution of perturbation equations of total matter fluid and individual fluids, we provide the following equations
\begin{eqnarray}
&\Delta_{m}=\frac{\mu_{d}}{\mu_{d}+\mu_{r}}\Delta_{d}+\frac{\mu_{r}}{\mu_{d}+\mu_{r}}\Delta_{r}\; , \label{dmdecomposed1}\\
&S_{dr}=\frac{\mu_{d}}{h_{d}}\Delta_{d}-\frac{\mu_{r}}{h_{r}}\Delta_{r}=\Delta_{d}-\frac{3}{4}\Delta_{r}\; .\label{Sdrdecomposed1} 
\end{eqnarray}
Because of that, we can write
\begin{equation}
\dot{S}_{dr}=\dot{\Delta}_{d}-\frac{3}{4}\dot{\Delta}_{r}\; .\label{dotSdrdecomposed1} 
\end{equation}
\subsection{Short-wavelength limits}
The short-wavelength modes are the modes for which the values of wave number $k$ are large.
For the short-wavelength limits, $\lambda$ 
is constrained with an upper boundary by the Jeans length for radiation perfect fluid $\lambda_{J}$, see \cite{amare4}.
Jean's length is defined  \cite{jeans1902stability,sahni1995approximation,alcubierre2015cosmological} as
$
\lambda_{J}=c_{s}\sqrt{\frac{\pi}{G\mu}}, 
$
where $c_{s}$ is the sound speed, $G$ is gravitational constant and $\mu$ is the energy density of baryonic matter. For decoupling epoch
(decoupling between radiation and matter), one has approximated value of Jeans's length as $\lambda_{J}=1.6 Mpc$. 
In this range of wavelength, one can choose 
to study different epochs of the considered system of radiation-dust mixture. 
The most studied epochs in the literatures are radiation-dominated epoch
and dust-dominated epoch (see more detail in \cite{amare4} and references herein).
Let us focus our interests on radiation-dominated epoch where the interaction between
component fluids is neglected.

\subsubsection{Radiation-dominated epoch}
In this context of radiation dominated over dust component, we assume the homogeneity of radiation energy density with flat universe $K=0$. These assumption results in having
$\Delta_{r}\approx 0$. 
Therefore, the evolution equations governing this system are 
\begin{eqnarray}
& \dot{\Delta}^{k}_{d}=-Z^{k}+\theta c^{2}_{sd}\Delta^{k}_{d}+\frac{k^{2}}{a}V^{k}_{d}
-\frac{\theta}{h}\Big(c^{2}_{s}\mu \Delta^{k}+p\varepsilon^{k}\Big)\; ,\\
&\begin{split}
&\dot{Z}^{k}=\Big(\frac{\dot{\phi}}{\phi+1}-\frac{2\theta}{3}\Big)Z^{k}+\frac{3k^{2}c^{2}_{s}}{4a^{2}}\Delta^{k}_{d}
+\frac{k^{2}}{4a^{2}}\varepsilon^{k}
+\frac{k^{2}}{a^{2}(\phi+1)}\Phi^{k}\\
&+\Big[\frac{1}{2\phi'}+\frac{2\mu_{r}-f-2\theta\dot{\phi}}{2(\phi+1)^{2}}+\frac{k^{2}\phi'' R}{a^{2}\phi'(\phi+1)}
-\frac{k^{2}\phi' R}{a^{2}(\phi+1)^{2}}\Big]\Phi^{k}
+\Big[\frac{3(\phi+1)c^{2}_{s}-4}{4(\phi+1)}\mu_{r}\\
&-\frac{3c^{2}_{s}}{4}\Big(-\frac{1}{3}\theta^{2}+\frac{f}{2(\phi+1)}+\frac{\theta\dot{\phi}}{\phi+1}
+\frac{k^{2}\phi'}{a^{2}(\phi+1)}R\Big)\Big]\Delta^{k}_{d}\\
&-4\Big[-\frac{1}{3}\theta^{2}-\frac{\mu_{r}}{\phi+1}+\frac{f}{2(\phi+1)}+\frac{\theta\dot{\phi}}{\phi+1}
+\frac{k^{2}\phi'}{a^{2}(\phi+1)}R\Big]\varepsilon^{k}+\frac{\theta}{\phi+1}\Psi^{k}\;,
\end{split}\\
&\dot{\Phi}^{k}=\Psi^{k}-\Big(\frac{c^{2}_{s}\dot{\phi}}{(w+1)}\frac{\mu_{d}}{(\mu_{d}+\mu_{r})}
+\frac{w\dot{\phi}}{w+1}\Big)\Delta^{k}_{d}\; ,\\
&
\begin{split}
 \dot{\Psi}^{k}&=\frac{\ddot{\phi}'}{\phi'}\Phi^{k}-\Big(\frac{c^{2}_{s}\ddot{\phi}}{(w+1)}\frac{\mu_{d}}{(\mu_{d}
 +\mu_{r})}+\frac{w\ddot{\phi}}{(w+1)}\Big)\Delta^{k}_{d}\; ,
\end{split}\\
&
\dot{V}^{k}_{d}-(3c^{2}_{sd}-1)\frac{\theta}{3}V^{k}_{d}=\frac{1}{ahh_{d}}\Big(-hc^{2}_{sd}\mu_{d}\Delta^{k}_{d}
+h_{d}(c^{2}_{s}\mu \Delta^{k}+p\varepsilon^{k})\Big)\; ,\\
&
\begin{split}
 \dot{V}^{k}_{dr}&=-(c^{2}_{sd}-c^{2}_{sr})\theta V^{k}_{d} -(3c^{2}_{sr}-1)\frac{\theta}{3}V^{k}_{dr}
 -\frac{(c^{2}_{sd}-c^{2}_{sr})}{a}\Delta^{k}_{d}-\frac{c^{2}_{sr}}{a}\Delta^{k}_{d}\; ,
\end{split}\\
& \dot{S}^{k}_{dr}=\frac{k^{2}}{a}V^{k}_{dr}\; .
\end{eqnarray}
The matter energy density and entropy become
\begin{eqnarray}
&&\Delta^{k}_{m}=\frac{\mu_{d}}{\mu_{d}+\mu_{r}}\Delta^{k}_{d}\; \;\;\text{and}\;\; S^{k}_{dr}=\Delta^{k}_{d}\; , 
\end{eqnarray}
so that the evolution equations become
\begin{eqnarray}
&&\dot{S}^{k}_{dr}=\dot{\Delta}^{k}_{d}\; ,\\
&&\varepsilon^{k}=-\frac{4\mu_{d}}{(3\mu_{d}+4\mu_{r})}(c^{2}_{sd}-c^{2}_{sr})\Delta^{k}_{d}\; ,\label{varepsilonkkkkk} \\
&&\dot{\varepsilon}^{k}=-4\theta c^{2}_{z}c^{2}_{s}\Delta_{d}-4c^{2}_{z}\dot{\Delta}^{k}_{d}\; .
\end{eqnarray}
The following assumption can help us to reduce the above equations once more. 
From the homogeneity of radiation energy density as background, one has the following
approximation \cite{amare4}
\begin{eqnarray}
&&\Delta_{r}<< \Delta_{d}\; . 
\end{eqnarray}
This results in 
\begin{eqnarray}
c^{2}_{s}\mu\Delta^{k}_{m}+p\varepsilon^{k}=\frac{1}{3}\mu_{r}\Delta^{k}_{r}\approx 0\; .\label{assumption11}
\end{eqnarray}
Therefore, our equations are reduced to
\begin{eqnarray}
 & \dot{\Delta}^{k}_{d}=-Z^{k}+\frac{k^{2}}{a}V^{k}_{d}\; ,\label{Deltakdradiationdominated1}\\
&\small
\begin{split}
&\dot{Z}^{k}=\Big(\frac{\dot{\phi}}{\phi+1}-\frac{2\theta}{3}\Big)Z^{k}+\frac{3k^{2}c^{2}_{s}}{4a^{2}}\Delta^{k}_{d}
+\frac{k^{2}}{4a^{2}}\frac{4\mu_{d}c^{2}_{sr}}{(3\mu_{d}+4\mu_{r})}
\Delta^{k}_{d}
+\frac{k^{2}}{a^{2}(\phi+1)}\Phi^{k}\\
&+\Big[\frac{1}{2\phi'}+\frac{2\mu_{r}-f-2\theta\dot{\phi}}{2(\phi+1)^{2}}+\frac{k^{2}\phi'' R}{a^{2}\phi'(\phi+1)}
-\frac{k^{2}\phi' R}{a^{2}(\phi+1)^{2}}\Big]\Phi^{k}
+\Big[\frac{3(\phi+1)c^{2}_{s}-4}{4(\phi+1)}\mu_{r}\\
&-\frac{3c^{2}_{s}}{4}\Big(-\frac{1}{3}\theta^{2}+\frac{f}{2(\phi+1)}+\frac{\theta\dot{\phi}}{\phi+1}
+\frac{k^{2}\phi'}{a^{2}(\phi+1)}R\Big)\Big]\Delta^{k}_{d}
-4\Big[-\frac{1}{3}\theta^{2}\\
&-\frac{\mu_{r}}{\phi+1}+\frac{f}{2(\phi+1)}
+\frac{\theta\dot{\phi}}{\phi+1}
+\frac{k^{2}\phi'}{a^{2}(\phi+1)}R\Big]\frac{4\mu_{d} c^{2}_{sr}}{(3\mu_{d}+4\mu_{r})}\Delta^{k}_{d}+\frac{\theta}{\phi+1}\Psi^{k}\;,
\end{split}\\
&\dot{\Phi}^{k}=\Psi^{k}\; \;\;\text{and}\;\; \dot{\Psi}^{k}=\frac{\ddot{\phi}'}{\phi'}\Phi^{k}\; ,\label{Psikdradiationdominated1kk}\\
 &\dot{V}^{k}_{d}+\frac{\theta}{3}V^{k}_{d}=0\; , \dot{V}^{k}_{dr}=\frac{\theta}{3}V^{k}_{dr}\; \;\;\text{and}\;\; \dot{S}^{k}_{dr}=\frac{k^{2}}{a}V^{k}_{dr}\; ,
\end{eqnarray}
where we have considered that $c^{2}_{sd}=0$ and Eq. \eqref{varepsilonkkkkk} is used. Here, we can set the direction of the unit velocity vector $u^{d}_{a}$ of the dust to be in the same direction as that of the total matter fluid.
This implies that we have a vanishing relative velocity $V^{d}_{a}$ according to Eq. \eqref{relativevelocitVi}. Thus we have
\begin{equation}
V^{k}_{d}=0\; , \label{uaequaluE1}
\end{equation}
so that the second-order equation in $\Delta^{k}_{d}$ becomes
\begin{equation}
\begin{split}
&\ddot{\Delta}^{k}_{d}-\Big(\frac{\dot{\phi}}{\phi+1}-\frac{2\theta}{3}\Big)\dot{\Delta}^{k}_{d}+\Big[\frac{k^{2}}{a^{2}(\phi+1)}+\frac{1}{2\phi'}+\frac{2\mu_{r}-f-2\theta\dot{\phi}}{2(\phi+1)^{2}}\\
&+\frac{k^{2}\phi'' R}{a^{2}\phi'(\phi+1)}
-\frac{k^{2}\phi' R}{a^{2}(\phi+1)^{2}}\Big]\Phi^{k}
+\Big[\frac{k^{2}}{4a^{2}}
+\frac{k^{2}}{4a^{2}}\frac{4\mu_{d}}{3(3\mu_{d}+4\mu_{r})}+\frac{(\phi+1)-4}{4(\phi+1)}\mu_{r}\\
&-\frac{1}{4}\Big(-\frac{1}{3}\theta^{2}+\frac{f}{2(\phi+1)}+\frac{\theta\dot{\phi}}{\phi+1}
+\frac{k^{2}\phi'}{a^{2}(\phi+1)}R\Big)\Big]\Delta^{k}_{d}
-4\Big[-\frac{1}{3}\theta^{2}-\frac{\mu_{r}}{\phi+1}\\
&+\frac{f}{2(\phi+1)} +\frac{\theta\dot{\phi}}{\phi+1}
+\frac{k^{2}\phi'}{a^{2}(\phi+1)}R\Big]\frac{4\mu_{d}}{3(3\mu_{d}+4\mu_{r})}\Delta^{k}_{d}+\frac{\theta}{\phi+1}\dot{\Phi}^{k}=0\;,
\end{split}
\end{equation}
where $c^{2}_{sr}=\frac{1}{3}$.
Once again, since we are dealing with the epoch of radiation dominating over dust, we also assume that the energy density of radiation
is much larger than that of dust, that is
$
\mu_{d}<<\mu_{r}, 
$
so that the leading equation becomes
\begin{equation}
\begin{split}
&\ddot{\Delta}^{k}_{d}-\Big(\frac{\dot{\phi}}{\phi+1}-\frac{2\theta}{3}\Big)\dot{\Delta}^{k}_{d}+\Big[\frac{k^{2}}{a^{2}(\phi+1)}+\frac{1}{2\phi'}+\frac{2\mu_{r}-f-2\theta\dot{\phi}}{2(\phi+1)^{2}}\\
&+\frac{k^{2}\phi'' R}{a^{2}\phi'(\phi+1)}
-\frac{k^{2}\phi' R}{a^{2}(\phi+1)^{2}}\Big]\Phi^{k}
+\Big[\frac{k^{2}}{4a^{2}}
+\frac{k^{2}}{12a^{2}}\frac{\mu_{d}}{\mu_{r}}+\frac{(\phi+1)-4}{4(\phi+1)}\mu_{r}\\
&-\frac{1}{4}\Big(-\frac{1}{3}\theta^{2}+\frac{f}{2(\phi+1)}+\frac{\theta\dot{\phi}}{\phi+1}
+\frac{k^{2}\phi'}{a^{2}(\phi+1)}R\Big)\Big]\Delta^{k}_{d}
-\frac{4}{3}\Big[-\frac{1}{3}\theta^{2}-\frac{\mu_{r}}{\phi+1}\\
&+\frac{f}{2(\phi+1)} +\frac{\theta\dot{\phi}}{\phi+1}
+\frac{k^{2}\phi'}{a^{2}(\phi+1)}R\Big]\frac{\mu_{d}}{\mu_{r}}\Delta^{k}_{d}+\frac{\theta}{\phi+1}\dot{\Phi}^{k}=0\;.
\end{split}
\end{equation}
We  approximate once more that the product of the dust energy density perturbation $\Delta^{k}_{d}$ and  dust energy density $\mu_{d}$
are small enough so that we neglect $\mu_{d}\Delta^{k}_{d}/\mu_{r}$ over the other terms. This assumption has also been used in many literatures, see the work done in \cite{amare4} for example.
Thus, we have an approximated equation as
\begin{equation}
\begin{split}
&\ddot{\Delta}^{k}_{d}-\Big(\frac{\dot{\phi}}{\phi+1}-\frac{2\theta}{3}\Big)\dot{\Delta}^{k}_{d}+\Big[\frac{k^{2}}{a^{2}(\phi+1)}+\frac{1}{2\phi'}+\frac{2\mu_{r}-f-2\theta\dot{\phi}}{2(\phi+1)^{2}}\\
&+\frac{k^{2}\phi'' R}{a^{2}\phi'(\phi+1)}
-\frac{k^{2}\phi' R}{a^{2}(\phi+1)^{2}}\Big]\Phi^{k}
+\Big[\frac{k^{2}}{4a^{2}}
+\frac{(\phi+1)-4}{4(\phi+1)}\mu_{r}\\
&-\frac{1}{4}\Big(-\frac{1}{3}\theta^{2}+\frac{f}{2(\phi+1)}+\frac{\theta\dot{\phi}}{\phi+1}
+\frac{k^{2}\phi'}{a^{2}(\phi+1)}R\Big)\Big]\Delta^{k}_{d}+\frac{\theta}{\phi+1}\dot{\Phi}^{k}=0\;.
\end{split}
\end{equation}
The second-order equation in $\Phi^{k}$ is obtained from Eq. \eqref{Psikdradiationdominated1kk} to be
\begin{equation}
\ddot{\Phi}^{k}-\frac{\ddot{\phi}'}{\phi'}\Phi^{k}=0\; . \label{eqbeforequasiPhi}
\end{equation}
In GR limiting case, one has \cite{amare4}
\begin{equation}
\ddot{\Delta}^{k}_{d}+\frac{2\theta}{3}\dot{\Delta}^{k}_{d}=0\;,\;\;\text{and}\;\;
\Phi^{k}=0\;.
\end{equation}
In \cite{amare4}, the quasi-static analysis is done where they have neglected time derivatives of curvature terms in their analysis. In our analysis, we can still
neglect time derivatives of the scalar field perturbations $\ddot{\Phi}^{k}$ and $\dot{\Phi}^{k}$, with the understanding that the scalar field perturbations evolve slowly with time in comparison to the matter perturbations.
From Eq. \eqref{eqbeforequasiPhi}, one can easily see how the quantity $\Phi^{k}$ vanishes as a consequence of quasi-static assumption.
We therefore have the reduced equation as
\begin{eqnarray}
\begin{split}
&\ddot{\Delta}^{k}_{d}-\Big(\frac{\dot{\phi}}{\phi+1}-\frac{2\theta}{3}\Big)\dot{\Delta}^{k}_{d}
+\Big[\frac{k^{2}}{4a^{2}}
+\frac{(\phi+1)-4}{4(\phi+1)}\mu_{r}\\
&-\frac{1}{4}\Big(-\frac{1}{3}\theta^{2}+\frac{f}{2(\phi+1)}+\frac{\theta\dot{\phi}}{\phi+1}
+\frac{k^{2}\phi'}{a^{2}(\phi+1)}R\Big)\Big]\Delta^{k}_{d}=0\;.\label{radiationquasi1}
\end{split}
\end{eqnarray}
\noindent To easily analyse the behavior of the energy density perturbations, we define $\delta(z)$ as
\begin{eqnarray}
&\delta (z)\equiv \frac{\Delta^{k}_{m}(z)}{\Delta_{in}}\;,
\end{eqnarray}
where $\Delta_{in}$ is the initial value to be chosen.
We consider the model of  power law $f(R)$ gravity studied in \cite{abebe2013large} given as
\begin{equation}
f(R)=\beta H^{2}_{0}\left(\frac{R}{H^{2}_{0}}\right)^{n}\;,
\end{equation}
where $H_{0}$ is the present value of the Hubble parameter, $\beta$ and $n$ are model parameters. For $R^{n}$ models \cite{barrow1983stability,barrow1988inflation}, from 
dynamical system analysis of stable points, it was found that
we can have a particular orbit in the phase space with solution of scale factor given as \cite{26,clifton2005power}
\begin{equation*}
a=a_{*}\Big(\frac{t}{t_{*}}\Big)^{\frac{2n}{3(1+\omega)}}\; ,
\end{equation*}
here $a_{*}$ is the scale factor at the time $t_{*}$ where  energy densities of the scalar field and that of matter were equal. In most cases $a_{*}$ 
and $t_{*}$ are normalized to unit \cite{amare4}, so that one writes the scale factor as
\begin{equation}
a=t^{\frac{2n}{3(1+\omega)}}\; . \label{scalefactort}
\end{equation}
We will also use the scale factor as function of redshift as
\begin{equation}
a=\frac{1}{1+z}\;.
\end{equation}
In the redshift space, eq. \eqref{radiationquasi1} is
\small{
\begin{eqnarray}
\begin{split}
&\Delta''_{d} +
 4\Big[ \frac{(1+z) ^{-3}}{2}+\frac{(1+z)^{\frac {-2-3n}{n}}}{2}
 +\frac{2\beta (n-1)A^{n-1}(1+z) ^{-\frac {2n-11}{2n}}}{ \beta\,nA^{n-1}(1+z)^{
3n-3}+1}+( 1+z) ^{\frac{-2-4n}{n}} \Big] 
 \frac{\Delta'_{d}}{n}\\
 & +4
\frac {\Delta_{d} }{{n}^{2}} \Big[ {\frac {\pi^{2}}{{
\lambda}^{2}}}+\frac{n\Omega_{r} \Big( 3\,\beta\,n(H_{0} A) ^{2(n-1)}(1+z)^{3n-3}-9 \Big)(1+z)^{\frac{2}{n}}}{4\beta n (H_{0} A) ^{2(n-1)}(1+z)^{3n-3}+4} 
+\frac{3n^{2}}{16}\sqrt [n]{1+z}\\
&-\frac{\frac{\beta}{4} H_{0}^{2}(1+z) ^{3}A^{n}}{2\,\beta\,n H_{0}^{2} A^{n-1}(1+z)^{\frac {3(n-1)}{n}}+2 }+\frac{3}{4}\frac{\beta\,{n}^{2} (n-1) A^{n-1}(1+z)^{-\frac {3(-2\,n+1)}{2n}}(1+z)^{\frac{2}{n}} }{ \beta\,nA^{n-1}(1+z)^{3n-3}+1 }\\
& -\frac{\frac {
{\pi}^{2}\beta\,n \left( n-1 \right) }{{\lambda}^{2}} A ^{n-1} \left( 1+z \right) ^{3
\,{\frac {n-1}{n}}}}{ \beta\,n A^{n-1} \left( 1+z \right) ^{3\,{\frac {n-1}{n}}
}+1} \Big]\frac{1}{(1+z) ^{{\frac {4+4\,
n}{n}}}}=0\;,
  \label{radiationquasi1bb}
 \end{split}
\end{eqnarray}}
\noindent where $A=\frac{4n^{2}-4n}{H_{0}^{2}}$ and the primes $''$ indicate derivatives with respect to redshift $z$, however, the primes present in the other equations that have time derivatives stand for derivatives with respect to the Ricci scalar. 
\begin{figure}[h!]
\centering
\includegraphics[scale=0.32]{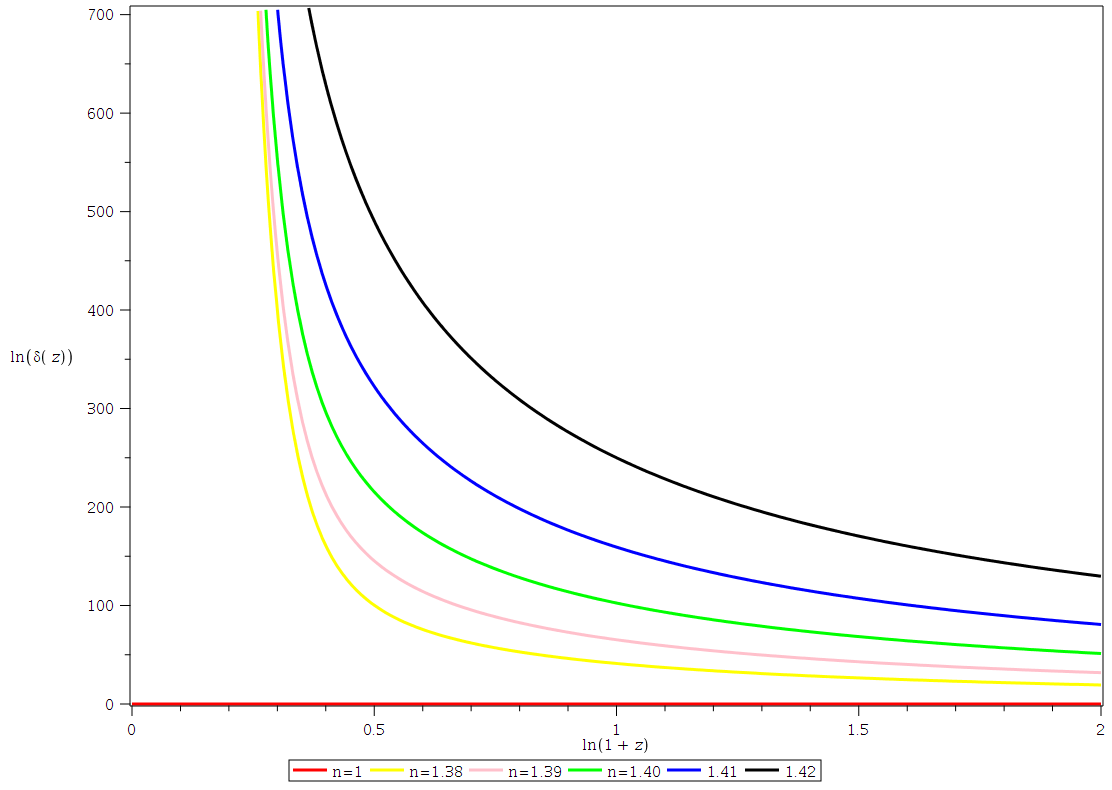}
\caption{Plot of dust energy density fluctuations as solution to Eq. \eqref{radiationquasi1bb} for different values of $n$, $\beta=10^{-10}$, $\lambda=1.5$Mpc, $\Omega_{r}=10^{-4}$ and $H_{0}=73$km/s/Mpc.}
\label{Figmultishortradiationk45lamdachanging3}
\end{figure}
\noindent In Fig. \ref{Figmultishortradiationk45lamdachanging3},
we have presented the energy density fluctuation of dust fluid for radiation dominated epoch. We have fixed $\lambda$ to be less than the Jean's wavelength $\lambda_{J}$ as $\lambda=1.5Mpc$. The initial value conditions can be considered being small enough. For instance the work done in \cite{abebe2013large} considered $\Delta_{d}(z_{0})=10^{-5}$ and $\Delta'_{d}(z_{0})=0$ for $z_{0}=2000$, we will more or less use the initial conditions slightly different from those presented in \cite{abebe2013large}. In Fig. \ref{Figmultishortradiationk45lamdachanging3}, we have used  $\Delta_{d}(z_{0})=10^{-9}$ and $\Delta'_{d}(z_{0})=0$ for $z_{0}=2000$. We considered value of $n$ to be in the range of $1.3<n<1.5$ \cite{abebe2013large,26}. It is clear that the energy density fluctuation decreases with the increase of redshift. This behavior of dust perturbations that evolved from the radiation background do not involve the influence of the perturbations due to the scalar field fluid. This is the consequence of the assumption of the quasi-static approximation made in this subsection. 
\subsubsection{Dust-dominated epoch}
In the dust dominated epoch, we assume that the universe is filled with pressure-less dust together with radiation. But this mixture are such that
the dust dominates the system. With this assumption, we consider that this background system together with scalar field fluid form an
homogeneous background of flat FLRW type. Therefore, we are treating the behavior of matter perturbations. 
That is, the dust perturbations arising in the that system. The energy density of radiation is much smaller than energy density
of dust in this dust dominated epoch,
\begin{equation}
 \mu_{r}<<\mu_{d}\; .
\end{equation}
Due to this assumption, we also assume that the perturbations due to radiation energy density are small enough compared to 
the perturbations generated from dust, that
\begin{equation}
\Delta_{r}<< \Delta_{d}\; . 
\end{equation}
Due to this assumption, and due to the earlier assumption that radiation is homogeneous anyway, we consider that $\Delta_{r}\approx 0$.
We are motivated to have such considerations due to the fact that, later, that is, after radiation dominance in the evolution of the 
universe, the favor can be given to the dominance of pressure-less dust to produce inhomogeneities in the universe.
With the above assumption, the evolution equations governing this system are
\begin{equation}
\begin{split}
 \dot{\Delta}^{k}_{d}&=-Z^{k}+\frac{k^{2}}{a}V^{k}_{d}\; ,\label{dotDeltammulti1dustdom}
\end{split} 
\end{equation}
where we have used the same assumption as in Eq. \eqref{assumption11}, $c^{2}_{sd}=c^{2}_{s}=0$ and $w=w_{d}=0,$ and
\begin{eqnarray}
 &&
\begin{split}
\dot{Z}^{k}=&\Big(\frac{\dot{\phi}}{\phi+1}-\frac{2\theta}{3}\Big)Z^{k}-\frac{\mu_{d}}{(\phi+1)}\Delta^{k}_{d}
+\Big[\frac{1}{2\phi'}+\frac{\mu_{d}}{(\phi+1)^{2}}-\frac{f}{2(\phi+1)^{2}}\\
&-\frac{\theta\dot{\phi}}{(\phi+1)^{2}}
+\frac{k^{2}\phi''R}{a^{2}\phi'(\phi+1)}
-\frac{k^{2}\phi'R}{a^{2}(\phi+1)^{2}}+\frac{k^{2}}{a^{2}(\phi+1)}\Big]\Phi^{k}+\frac{\theta}{\phi+1}\Psi^{k}\; ,
\label{dotZmulti12111dustdom}
\end{split}\\
&&\dot{\Phi}^{k}=\Psi^{k}\; ,\;\;\text{and}\;\; \dot{\Psi}^{k}=\frac{\ddot{\phi}'}{\phi'}\Phi^{k}\; ,\label{dotPsimulti12111dustdom}\\
&&\dot{V}^{k}_{d}+\frac{\theta}{3}V^{k}_{d}=0\; , \;\; \dot{V}^{k}_{dr}= \frac{\theta}{3}V^{k}_{dr}\; ,\;\;\text{and}\;\; \dot{S}^{k}_{dr}=\frac{k^{2}}{a}V^{k}_{dr}\; . \label{dotSilmulti1harmonic111dustdom}
\end{eqnarray}
The scalar gradient variable for matter and entropy become
\begin{eqnarray}
&&\Delta_{m}=\frac{\mu_{d}}{\mu_{d}+\mu_{r}}\Delta_{d}+\frac{\mu_{r}}{\mu_{d}+\mu_{r}}\Delta_{r}\approx \Delta_{d}\;
, \label{dmdecomposed1dustdom}\\
&&S_{dr}=\frac{\mu_{d}}{h_{d}}\Delta_{d}-\frac{\mu_{r}}{h_{r}}\Delta_{r}=\Delta_{d}\; .\label{Sdrdecomposed1dustdom} 
\end{eqnarray}
The evolution equation is
\begin{equation}
\dot{S}_{dr}=\dot{\Delta}_{d}\; .\label{dotSdrdecomposed1dustdom} 
\end{equation}
Then one has
\begin{eqnarray}
\begin{split}
 &\ddot{\Delta}^{k}_{d} -\Big(\frac{\dot{\phi}}{\phi+1}-\frac{2\theta}{3}\Big)\dot{\Delta}^{k}_{d} 
 -\frac{\mu_{d}}{(\phi+1)}\Delta^{k}_{d}=0\; ,
 \label{ddotafterquasi122}
\end{split} 
\end{eqnarray}
where the approximation of quasi-static is assumed such that 
$\dot{\Phi}^{k}=\Psi^{k}\approx 0\; ,$ and 
$ \dot{\Psi}^{k}=\frac{\ddot{\phi}'}{\phi'}\Phi^{k}\approx 0$.
In the GR limit, Eq. \eqref{ddotafterquasi122} reduces to \cite{amare4}
\begin{eqnarray}
\begin{split}
 &\ddot{\Delta}^{k}_{d}+\frac{2\theta}{3}\dot{\Delta}^{k}_{d} 
 -\mu_{d}\Delta^{k}_{d}=0\; ,
 \label{GRddotafterquasi2}
\end{split} 
\end{eqnarray}
In the redshift space, we have Eq. \eqref{ddotafterquasi122} as
\begin{equation}
\begin{split}
&\frac{4n^{2}}{9}(1+z)^{\frac {3+2\,n}{n}}\Delta''_{d} + \Big[ \frac{2n}{3}(1+
z)^{\frac {3+n}{n}}+\frac{4n^{2}}{9}(1+z) ^{
\frac {3+n}{n}}\\
&+\frac{6\times \big(\frac{4}{3} \big)^{n}\beta(n-1) H_{0}^{2} ( 1+z)^{\frac{6n+3}{2n}} B ^{n}}{3^{-n+1}\beta H_{0}^{2} (1+z) ^{\frac{3n-3}{n}}{4}^{n} B ^{n}+16\,n-12} +4n( 1+z) ^{\frac{3}{2n}} \Big]\Delta'_{d}  \\
&\frac{-8n\Omega_{d}( 4n-3)(1+z) ^{\frac{3}{2n}}}{ {3}^{-n+1}\beta H_{0}^{2}  ( 1+z) ^{\frac{3n-3}{n}} {4}^{n} B ^{n}+16\,n-12 } \Delta_{d}=0 \;, \label{ddotafterquasi211}
\end{split}
\end{equation}
where $B=\frac {n \Big( 4\,n-3
 \Big) }{H_{0}^{2}}$.
\begin{figure}[h!]
\centering
\includegraphics[scale=0.3]{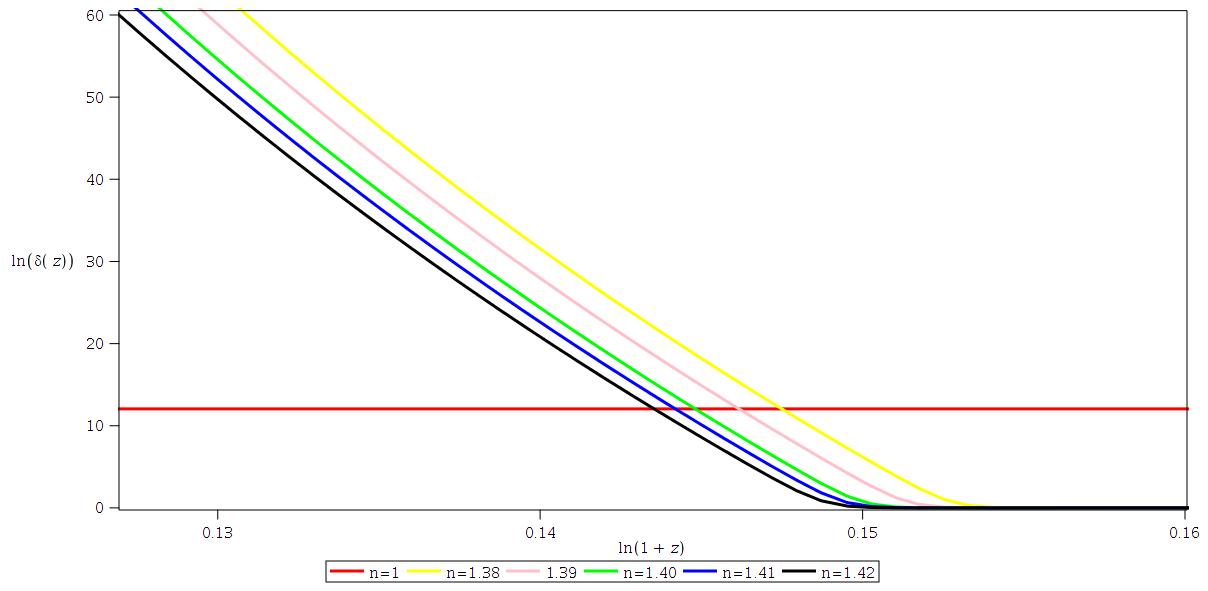}
\caption{Plot of dust energy density fluctuations for dust-dominated in the shortwavelength limit, solution of Eq. \eqref{ddotafterquasi211} for $\Omega_{d}=0.32$ and $H_{0}=73$km/s/Mpc, $\beta=0.005$\;.}
\label{short_dust1a}
\end{figure}

\noindent For the dust-dominated situation, see Fig. \ref{short_dust1a}, the dust perturbations from the ST theory look bigger in lower redshifts but became lower than that of GR theory ($n=1$ and $\beta=1$, see red ligne in Fig. \ref{short_dust1a}) at high redshift. However, one could notice that the solution presented here together with dust-dominated medium in the background has no influence of perturbation due to the scalar field as a consequence of the assumption of quasistatic approximation considered.
 We used the initial value conditions to $\Delta_{d}(z_{0})=10^{-3}$ and $\Delta'_{d}(z_{0})=0$. The choice was made following the work done in \cite{abebe2013large}.


\noindent For the general concerns, one can notice that some of the gradient variables are absent in the final equations due to several reasons. For instance, the entropy gradient variable $\varepsilon$ (as well as the relative energy density gradient $S_{ij}$) has been written in terms of the matter gradient variables $\Delta_{d}$ and $\Delta_{r}$ of dust and radiation respectively. The gradient variable due to relative velocity is washed away due to the considered assumption that the direction of unit vector of the individual fluid (dust/radiation) is set to be in the same direction as that of the total fluid. Also depending on the dominating fluid, the fluid energy densities have been approximated either like $\mu_{d}<<\mu_{r}$ or the other way around. These considerations have dramatically simplified our equations together with the use of harmonic decomposition and quasi-static approximation in the short-wavelength limit. The observational values used in the calculations are taken from \cite{ade2015planck}.

\section{Conclusions and discussions}\label{conclusion}
In this paper, we treated linear covariant perturbations of a flat FLRW spacetime background using the $1+3$ covariant and gauge-invariant perturbations approach.
We use the equivalence between Brans-Dicke type scalar-tensor theory and $f(R)$ theory to define gradient variables and their evolution equations
based on the extra degree of freedom considered as a scalar field. 
The medium under treatment is a scalar field-standard matter non-interacting
multi-fluid system where after
scalar and harmonic decompositions the focus specifically went to radiation-dust system. We only considered power law $f(R)$ model for the current study. 
The short-wavelength limit is considered. Two subsystems are treated namely radiation- and dust-dominated respectively. The consideration of quasi-static approximation is done where in the radiation-dominated epoch, it has been observed that the
density fluctuations of the  dust energy density behave in such a way that they decrease with the increase in the redshift. The same inspection has been observed in the dust dominated epoch. The density fluctuation of dust energy density decreases with the increase of the redshift.
The range of the considered values of $n$ was taken from the work done for power-law $f(R)$ models by \cite{abebe2013large,26}, where it was shown that this range represents solutions of a stable orbit with a decelerated matter dominated solution and late-time power-law acceleration. Additionally, most of the initial values considered were adopted from the work done in \cite{abebe2013large}. 
 However, much work needs to be done especially in the consideration of long-wavelength regime where one needs to consider the adiabatic
initial conditions, i.e., vanishing relative entropy perturbations on large scales; other different types of $f(R)$ models to extensively study if the equivalence between $f(R)$ gravity the scalar-tensor theory holds at the linear perturbative level.  

\section*{Acknowledgments} 
JN and MM gratefully acknowledge financial support from the Swedish International Development Cooperation Agency (SIDA) 
through the International Science Program (ISP) to the University of Rwanda (Rwanda Astrophysics, Space and Climate Science Research Group) grant number RWA01.
AA  acknowledges that this work is based on the research supported in part by the National Research Foundation of South Africa (grant number 112131). The authors thank the anonymous referee for the critical points raised that led to an improved version of this manuscript.


 
%
 
\end{document}